\documentclass[AMA]{WileyASNA-v1}

\articletype{Review}%

\received{26 April 2016}
\revised{6 June 2016}
\accepted{6 June 2016}

\begin{document}

\title{Carl Wirtz' article from 1924 in Astronomische Nachrichten on the radial motions of  spiral nebulae}

\author[1]{Tom Richtler}

\address[1]{\orgdiv{Departamento de Astronom\'{\i}a}, \orgname{Universidad de Concepci\'on, Concepc\'ion},  \orgaddress{\state{} \country{Chile}}}


\corres{Tom Richtler \email{tom@astro-udec.cl}}

\authormark{}


\corres{Tom Richtler
\email{tom@astro-udec.cl}}

\presentaddress{Coliumo, Tom\'e, Chile}

\abstract{In the year 1924, a paper by Carl Wirtz appeared in ''Astronomische Nachrichten", entitled "De Sitter's cosmology and the radial motion of spiral galaxies". This paper and its author remained
largely unnoticed by the community, but it seems to be the first cosmological interpretation of the redshift of galaxies as a time dilation effect and the expansion of the Universe.  Edwin Hubble knew
Wirtz' publications  quite well.  
  The modern reader would find Wirtz' own understanding diffuse and contradictory in some aspects, but  that reflected the early literature on  nebulae, to which he himself
made   important contributions. 
The  100th anniversary provides a good opportunity to present an English transcription to the community, which can be found in the appendix.  This anniversary also provokes
to ask for the present status of cosmology which many authors see in a crisis. From an observational viewpoint it shall be  illustrated
that  until today there is no consistent/convincing understanding of how the Universe evolved. }

\maketitle

\section{Carl Wirtz and  the expansion of the Universe, the beginning of modern observational cosmology}
  
  \subsection{Carl Wirtz is not  prominent}
   The discovery of the expansion of the Universe marked the beginning of modern observational cosmology. How that happened was 
  rather  a process than a moment (e.g. \citealt{vandenbergh11, gron18, cervantes23}) and the prominent names are known to every astronomer and with Einstein even outside the astronomical community.
  Relativistic cosmology
  has been founded by Einstein himself and de Sitter, but de Sitter was also a key author for making Einstein's theory quickly known to the English speaking community \citep{desitter16a,desitter16b} (de Sitter wrote German articles as well). The same can be said for \citet{eddington23} and \citet{weyl22}.   

 Friedmann's fundamental papers on world models as general solutions of Einstein's field equations \citep{friedmann22,friedmann24} 
     were written in German  and were more or less ignored by the community for about a decade  (although Einstein knew them well) until 
  \citet{robertson33}   presented  his review of relativistic cosmology. Friedmann already introduced the beginning of he Universe and described pulsating
  Universes with a period of the order $10^{10}$ years. 
  
 On the observational side, the question who  discovered the expansion normally leads to the names of Lemaitre and Hubble. However, in the recent
 review article of \citet{cervantes23}, a paragraph is also devoted to Carl Wirtz,
 a name that   
is not known to every astronomer.  The low prominence the work of Carl Wirtz  enjoys in comparison with Lemaitre and Hubble, is inappropriate. 

 There are a few modern references to Wirtz in the English literature, most notably \citet{duerbeck89}, \citet{seitter99} and \citet{duerbeck02}, now also 
  \citet{cervantes23}. 
   There
are a few more in German in amateur journals, we  cite \citet{priester87}, \citet{duerbeck90a}, \citet{duerbeck90b}, and \citet{appenzeller09}.

In the year 1924, Wirtz published  in "Astronomische Nachrichten" a paper \citep{wirtz24} which has the characteristics to count as the starting point of modern observational cosmology (at the time of writing: 22 citations). We find its 100th anniversary a nice opportunity to present that short paper in an English translation to the
 community in the same journal. The translated article is presented in the Appendix.

 Basic biographical dates:  Carl Wirtz was born 1876 in Krefeld and died 1938 in Hamburg. He got his PhD 1899 at Bonn University, worked as assistant in Vienna until 1901, then
moved to Strasbourg,  where he got a professorship in 1909 and stayed there until 1919. Then he moved to Kiel as professor and was forced into early retirement 1937. 
In the following, the work of Carl Wirtz shall be briefly introduced in the context of cosmological quest in the early 20th century. 

\subsection{Earlier work of Carl Wirtz in the context of cosmology from 1915 -1923} 

It was Schwarzschild who introduced the concept of curvature of space in cosmology  \citep{schwarzschild00,stewart98}.
He estimated the minimal curvature radius of the Universe to be approximately
$10^8$ astronomical units by consideration of the maximal inhomogeneity of the distribution of stars. 
When \citet{einstein15} published his famous equation connecting gravity with the geometry of space-time, the Universe was still commonly understood as consisting mainly of a stellar system
populating a large volume, as Schwarzschild did. The distinction between Galactic and extragalactic objects had no reason. The distances and nature  of the spiral nebulae were unknown.

\citet{desitter17} published his world models as cosmological solutions to Einstein's equations.
De Sitter's Universe (his model B) was finite with a certain curvature radius R, static and empty. Time delation with distance r resulted
through  the metric tensor element $g_{44}$  which in the four-dimensional line element $ds^2  = ... + cos^2( r/R) c^2 dt^2$ effectively
gives a non-linear relation between redshift and distance.
de Sitter himself tried to identify  cosmological redshifts among various types of
stars and only had the Andromeda nebula, NGC 1068, and NGC 4696 as extragalactic objects. But Andromeda
is approaching and the other two galaxies show quite similar velocities about 1000 km/s, so the effect
of increasing time dilation with distance which he hoped to see, was not visible.

At about the same time \citet{wirtz18} wrote a paper on proper motions of 378 spiral nebulae.  That was a topic that
he had addressed earlier in several contributions. He used the  (spurious) proper motions  to derive  an apex 
of the solar motion quite similar to that from  stellar streams.  The mean proper motion for the spirals resulted to 0.027 "/year.  However, this time he also compiled 15 radial velocities
of spiral galaxies and noted the higher velocities with respect to the stellar system, but also that a
common velocity and a vertex direction is not a good representation. The introduction of a constant K
 independent
from the coordinates at the sphere led to a much better representation through the equation

\begin{equation}
\label{eq:airy}
v = V_x*cos \alpha ~ cos \delta + V_y*sin \alpha~ cos \delta + V_z*sin \delta + K
\end{equation}
v being the  velocity of an individual object at spherical coordinates $\alpha$ and $\delta$, and $V_x$,$V_y$,$V_z$ the velocity
components of the (negative) apex  motion of the observer. 

He got the best fit of the above relation for $V_x = -145$ km/s, $V_y = -268$ km/s, $V_z =  -762 $ km/s (i.e. $V = -820$ km/s),
 and $ K=+656$ km/s, and interpreted V as the radial correspondence to the proper motion. 
Then it is straightforward to derive a mean distance for the spiral nebulae. His result was 6.66 kpc and he concluded that the spiral nebulae
are located outside the Milky Way system. He then tried to interpret the "strange" constant K.  
 In his own words: {\it If we give this
value a literal interpretation, then it means that the system of spiral nebulae drives apart with respect to the
actual position of the solar system as the centre with a velocity of +656 km/s}. The paper concludes:  {\it Also
in the case of the nebulae  one expects that we hold single threads of a mesh in hands whose complete pattern
we still cannot disentangle.}

Four years later Wirtz published a paper entitled {\it On the statistics of the radial motions of spiral
nebulae and globular clusters} \citep{wirtz22}, again in "Astronomische Nachrichten". Strikingly, proper motions of the nebulae that had accompanied his scientific career for a long time,  were not mentioned at all. Apparently this chapter
had been closed with the previous paper. Now Wirtz lists only radial velocities of 29 spiral nebulae and
10 globular clusters. For the spiral nebulae, he finds a somewhat higher K-term of +887 km/s, meaning that the
entire system is moving away with this common radial velocity. He  now
considers the residuals to this common movements which he calls "errors" and understands them as the  peculiar
velocities of the spiral galaxies.
 For these velocities, he finds a  slight dependence
on Galactic latitude in the sense that the polar nebulae have higher velocities and on apparent magnitude, where the brightest
nebulae are approaching and the fainter ones receding. This dependence also appears with the angular diameters. 

Wirtz: {\it Of course,
one finds the analog relation also with the diameter of the nebulae, where small diameters belong to a positive change
of the distance, large diameters to a negative change. These statistical phenomena occur on top of the most striking and
principal instance that can be described as if the system of spiral nebulae was drifting apart relative to our position...
and the dependence on magnitude indicates that the nearest or the most massive spiral nebulae show a slower  outward motion 
than the more distant  or the less massive ones.}

Here appears for the first time the notion of a radial velocity dependent on distance, but disguised as a second order effect that is still  covered by the  imagination of a system of spiral galaxies moving away with a constant velocity. In modern language, Wirtz had to deal with the fact that many of his galaxies are not  in the Hubble-flow.

Wirtz' main observational work in Strasbourg was the photometry of nebulae, including globular clusters. This resulted in a catalogue of surface brightnesses, total brightnesses, and
angular diameters for 566 objects \citep{wirtz23}. Presumably, Wirtz saw   that apparent magnitudes and angular diameters are not well correlated and therefore did not use
magnitudes as distance indicators.  

\subsection{The role of Friedmann}
 
  Two publications of the mathematician, meteorologist, geophysicist, pilot 
and adventurer dominated the cosmological field for (at least!) 100 years.  
In a static Universe the old mystification of circular orbits may have survived. Planetary orbits must(!) be circular, elliptical
orbits are ugly. And the Universe cannot be variable! In this analogy, Friedmann's role  is the closest to Kepler's.
In an expanding Friedmann universe, time dilation results naturally as proportional to 1+z   and a linear velocity-distance relation is necessary
to maintain homogeneity.

In 1922 \citet{friedmann22} already introduced "non-stationary" world models (in German). There are no
 indications that Wirtz was aware of Friedmann's work, but  even Einstein needed some time  to see that
Friedmann's  solutions were indeed correct.  The most famous physicist of the epoch gives in a  brief note his blessings to the concept of an
expanding Universe and is ignored himself \citep{einstein23}! Perhaps the German language again was an obstacle
for the international community. Also Friedmann's last paper \citep{friedmann24} did not express interest in observational evidence
for his models.

\subsection{The work of  Silberstein, Oepik, Lundmark and Str\"omberg}

In the same year 1924 appeared the work of \citet{silberstein24}. Silberstein
 understood  de Sitter's time dilation as a relativistic Doppler effect and derived a relation between redshift and
curvature radius of the Universe. To determine this radius R, he used the radial velocities of 7 globular clusters  and got $R = 1.1\times10^7$ pc.
 \citet{silberstein29} still argued strongly for a value of $1.67\times 10^6$pc, meaning that in his de Sitter model,
the largest possible distance in the Universe would be  $1.3\times10^6$ pc, contradicting strongly  the
by then established extragalactic distances.

The extragalactic nature of the Andromeda nebula was by 1922 quite clear. \citet{oepik22}, by using the
rotational velocity and applying an early
version of the Tully-Fisher relation, got a distance of 500000 pc,  three years before \citet{hubble25} published in a much more famous paper the
 Cepheid distance of NGC 6822 of 214 000 pc, in Hubble's words {\it the first object definitely assigned to a region outside the galactic system}.

However, it is \citet{lundmark24a} who deserved the credit to have shown the first "Hubble-diagram", although a velocity-distance relation
is not recognisable in his graph.
As for Silberstein, his main objective was to measure the curvature radius of a  de Sitter Universe, and not the
nature of spiral galaxies.
For that he needed absolute distances for the spiral nebulae. He used the Andromeda nebulae as the reference  distance of
200 000 pc which he got from comparing Galactic novae with novae in Andromeda. He explicitly did not exclude a distance of 500 000 pc, as favoured by \citet{oepik22}.

 Lundmark adopted Silberstein's formula and interpretation, but with a much enlarged data set, ranging from nearby
stars over Cepheids and novae
to the sample of spiral galaxies which had been used also by Wirtz (he apparently got the spiral galaxy data personally from
Slipher, \citealt{lundmark24b}). Lundmark determined individual distances for his  spiral galaxies through the
comparison with Andromeda regarding brightnesses and angular diameters. Without an established distance scale, the individual
errors are of course large, particularly the three most distant galaxies (on Lundmark's scale)  NGC 278, NGC 1700, and
NGC 2681 show moderate velocities around 700 km/s. And many nebulae are, in modern terms, not in the Hubble flow, unfortunately also
Andromeda which is the reference. Lundmark also gives "mean distances" for his sample by the comparison
of the K-term with (spurious) proper motions. Although he labels the proper motion values as "illusory" and "only of use for giving
a upper limit to the motion", the mere fact that there is a common proper motion of spiral nebulae, clashed with the
concept of expansion. 

A third author who used Slipher's radial velocities in a cosmological
context was Gustaf Str\"omberg \citep{stroemberg25}.
He performed his analysis without the knowledge of Lundmarks results, but
expressed in the published paper his satisfaction about the agreement with
the main result, namely that there is no relation of redshift with distance.
Str\"omberg calculated the distances for spiral nebulae as $10^{0.2 m}$ where
the m are apparent magnitudes as given by \citet{wirtz23}! 
He only gives correlation coefficients.
The strongest correlation has been found with the angular distance
from the sun's apex. As one sees already from Lundmarks sample, the individual
distances based on magnitudes are simply too uncertain to uncover any
velocity-distance relation.
One may find
some irony in the fact that Wirtz' own measurements are used against his
findings.

Str\"omberg moreover understood the redshifts as caused by de Sitter's time dilation not
as indicating a real motion, which he considers to be  "ficticious".

\subsection{Wirtz' influence on Hubble}

To what degree Hubble was influenced by the reading of Wirtz' publications will remain elusive  in detail (see  \citealt{way13} for more comments on Hubble's work) , but 
there is no doubt that Hubble knew Wirtz' 1924 paper (and others) quite well.  The German language apparently was no obstacle to him. 
He cited regularly German papers, but firstly not Wirtz, particularly not in  \citet{hubble29} nor in \citet{hubble31}.   
However, Wirtz occupies a quite prominent place in Hubble's book "The realm of the nebulae" from 1936. In the chapter    
"The velocity-distance relation" Hubble describes the content of \citet{wirtz18} and \citet{wirtz22} in some detail (for the latter paper, he cites the year as 1921, not
1922). But he praises Wirtz mainly as the discoverer of the "K-term", the mean residual of radial velocities of the nebulae after correction for solar motion.  Hubble calls Wirtz even  "the leader in the field".  He qualifies the results of the 1924 paper  as
"suggestive rather than definitive". In conjunction
with the discussion of the work of \citet{lundmark24b} and \citet{stromberg25}, Hubble  concluded that by 1925, "the data did not establish a relation" (between distance and 
radial velocity) and so repeated Lundmark's and Str\"omberg's statements, probably without looking further into the details of their distance determinations. 
 Nevertheless, Hubble adds a footnote "Wirtz later published a stimulating popular presentation of the investigation and the implications of the
results ({\it Scientia}, 38, 303, 1925) in which he assumes that de Sitter's prediction has been verified."

The 100 years after Wirtz' paper have mainly seen an acceptance of Friedmann's understanding of the expansion of the Universe, culminating in the emergence
of the "standard cosmology" with Dark Matter and Dark Energy  as its pillars, simultaneously its threats. 

\section{Do we {\it really} understand the cosmos?}

 I shall  give a brief account of literature  with the emphasis on those works 
that represent  the main problematic issues in modern cosmology,  and that do not appear in otherwise exhaustive reviews of the history of
the Hubble constant (e.g. \citealt{cervantes23}).  The breath-taking progress in both observations and theory let any review to be quickly outdated (e.g.  \citealt{lopez17},  \citealt{amendola13}). 
This section shall demonstrate that the full understanding of galaxy redshifts is a process that until today is still going on.

\subsection{Standard cosmology and challenges}  
 The following sampling
 of literature highlights the search for observable expansion signatures and shall illustrate that the "standard cosmology" faces serious challenges.
  The section title plagiarises \citet{paddy17} who analyses  profoundly  cosmological fundamentals from the viewpoint of a theoretical
  physicist. As for the phenomenon of expansion, he asks "How come a cosmological arrow of time emerges from equations of motion which are
time-reversal invariant?" and suggests to search the answer in a quantum description of spacetime.  For humble astronomers, the directly observable (at least principally) properties of expansion,  most notably the value of the Hubble constant and the
  deceleration parameter, 
  formed  the core of the cosmological endeavour for a long time \citep{sandage70,sandage98,sandage99}.

  The spectacular successes in understanding the Cosmic Microwave Background, the distance scale and the large scale structure joined to formulate a "standard cosmology" where the total energy
  density of a Friedmann Universe is the sum of a baryonic part, Dark Matter, and Dark Energy (e.g. \citealt{planckcoll15}).
 The current price to be paid for successful fits of observed data of various kinds to a Friedmann cosmological scheme is that the major
 part of the 
 energy density remains unexplained. However, dark matter and dark energy may not exist at all (see sections \ref{sec:backreaction} and 
 \ref{sec:mond}).  There is hope    
that data from the  James Webb Space Telescope (JWST) and other big telescopes to come will lead to a better understanding in the
near future  (e.g. \citealt{lovyagin22}).

\subsection{Expansion signatures: time dilation and angular distances}
 Time dilation proportional to (1+z) as a signature of expansion has been first observed in the broadening of SNIa lightcurves \citep{goldhaber96,leibundgut96}, which later was confirmed by a larger set of SNIa of
 higher redshifts \citep{blondin08}. 
 
 Quasar variability was another high-redshift phenomenon for which time dilation could be expected. For a long time irregular variability covered any potential signal and  \citet{hawkins10}  firmly concluded
 that there is no time dilation.  This is contradicted by   \citet{lewis23} who used long-term photometry of a large sample of quasars and indeed found cosmological time dilation.
  
 Another signature of expansion  is the growth of angular sizes for high-redshift objects. Because nearby objects 
 become smaller with increasing distances, there must be  a maximum of angular distances in an expanding universe.  The lack of convincing rods has long impeded a proof of  the reality
 of this maximum
  \citep{melia18}, who also provide a history of this relation.  \citet{lovyagin22},  collecting galaxy sizes recently measured  with JWST (cautiously) conclude  that the expanding-Universe model fails and that
  the galaxy ages might be much larger than expected for a $\Lambda$CDM Universe.
   This might be indicative of an early "loitering" phase of the Universe, when expansion almost stalled, which simultaneously boosted structure formation \citep{sahni92}. 
  
\subsection{Accelerated expansion and backreaction}    
\label{sec:backreaction}
  The apparent acceleration  of the expansion had been worthy  a Nobel price \citep{riess98,perlmutter99,schmidt98} (together at the time of writing: 32038 citations). But see also \citet{yoshii95} who, by counting galaxies,  probed volumes instead of distances. Being
  less explicit, they leave it to the reader, who is familiar with Friedmann models, to see an accelerated
  expansion in their Fig.2 (at the time of writing: 41 citations),  and accordingly the existence of a cosmological constant or "Dark Energy" (e.g. \citealt{planckcoll15}).  This concept has been criticised by authors who pointed out the non-linearity of the Einstein-equation  and  that a mean density, simply averaged over large volumes as it appears in the Friedmann-equation, is not an adequate description in a general relativistic framework.
  In their view, the cosmological constant emerges from a comparison of "apples with oranges", from comparing a Universe without structure (the CMB) with a Universe with structure (today's Universe).
  The effect of structure formation on the dynamical fate of the Universe has been called "backreaction" \citep{wiltshire07,buchert08, buchert12,buchert23}. 
  
  The apparent tension between a local $H_0$ and a "CMB-$H_0$" provoked an intense discussion   \citep{kumar23,kroupa23}. Again, it may well be traced back to the tension between a Universe with structure and
a Universe without structure, whether  inhomogeneities  are global \citep{heinesen20}  or  local  \citep{mazurenko24}. 

\subsection{Dark Matter and MOND}  
\label{sec:mond}
  The other main ingredient of  Standard Cosmology is Dark Matter (from the huge literature we cite \citealt{bertone18}).  However, all searches for particle dark matter in terrestrial detectors 
 have been futile so far. Alternative concepts like Modified Newtonian Dynamics
(\citealt{milgrom83a,milgrom83b, milgrom83c, milgrom23b,bekenstein84, famaey12, kroupa12a, kroupa12b, kroupa15}) gained  weight over the last decades. With today's knowledge of galaxy dynamics, the existence of MONDian phenomenology among galaxies is now well
  confirmed (e.g. \citealt{lelli17,richtler24}). Weak lensing data show the constancy of circular rotation curves of isolated galaxies out to hundreds of kpc, beyond the expected virial radii of dark matter
  halos \citep{brouwer21,mistele24}.
  
   The long-standing allegation that MOND misses a theoretical basis, now appears  unfounded.  MOND can be understood as modified inertia rather than modified gravity
  \citep{milgrom22,milgrom23a}.
  \citet{bekenstein04} was the first to present a relativistic  theory with MOND as the non-relativistic limit, but that turned out to be incompatible with the equality of the speed of light and gravitational waves \citep{abbott17}. A new theory  \citep{skordis21,skordis22} respects this important observation and reproduces the CMB structure.
  
   One can also pave a direct way from the thermodynamics of horizons \citep{jacobson95},
  describing gravity as emerging from microscopic degrees of freedom  \citep{paddy15}, to a MOND-like gravity as worked out by \citet{hossenfelder17} and \citet{hossenfelder18}.   
  
  MOND also may be dominant for early structure formation \citep{sanders98, wittenburg23}, so backreaction effects are expected to be even stronger than in a standard Friedmann evolution.
  The influence  MOND or its underlying theory   on cosmology is only at the verge of being explored. It might dramatically change the current picture.

 \section{Final remarks: today's relevance of Carl Wirtz}
 
 The few numbers in the
table in Wirtz' paper may be seen as the first manifestation of one of the greatest  
discoveries in the history of physics. What the understanding
of the contemporary authors was, we can only guess.
 But what can be shown 
is that it needed decades for the astronomical community to find a proper
cosmological understanding of galaxy redshifts. It is very illustrative
 to scan through Appendix B of \citet{davis04} and see what famous names 
have not been free of misconceptions. 
 
 Following again  \citet{paddy17},     modern cosmology is characterised by a disproportion between {\it knowledge} that increased tremendously during the last decades, and {\it understanding} that
 still uses the framework given by Friedmann.    Wirtz and his fellow astronomers stand for the beginning of knowledge. Reading their papers  today  does not contribute to the understanding of cosmological expansion, but 
 might contribute to the understanding of the emergence of modern cosmology.

\section{acknowledgements}
T.R.  thanks Ralf-J\"urgen Dettmar for encouragement and discussion, furthermore the Astronomisches Institut der Ruhr-Universit\"at Bochum for hospitality.  Neil Nagar is thanked for critical reading 
and Sebastian Kamann and Pavel Kroupa for  literature advice and discussion.

\appendix


{ \fontfamily{qbk}\selectfont
\fontsize{11}{12}\selectfont 
\section{Carl Wirtz:  De Sitter's cosmology and the radial motions  of spiral galaxies}
The translation German - English is as literal as  is justifiable. The unit km should be read as km/s.\\

 {\large Astronomische Nachrichten 1924, 222, 21W}
\label{sec:paper}

The world of de Sitter is a spherical space-time world, a four-dimensional continuum of space and imaginary time that forms the surface
of a sphere in five dimensions. De Sitter's world is devoid of mass, all mass swept away to a zone that is not accessible for observations,
like a "mass-horizon" or a ring of peripheral matter, necessary to keep the emptiness within. The two limits that embrace the real world, are Einsteins world with a maximum
of matter and de Sitter's world.  W. de Sitter, MN 78.3, 1918;  A.S. Eddington, Math. Theory of Relativity, Cambridge 1923, Chap III,V; compare also the illustrative graphical representation
of de Sitter's world at  H. Weyl, Naturwiss.12 (1924), p.201. 

In de Sitter's world, system B, occurs a phenomenon  accessible to observations.  The natural processes are slowing down  with increasing distance from the origin of coordinates, so
do the natural clocks. In particular, atomic oscillations slow down. This decrease of  the oscillatory periods  can be observed: the spectral lines that come 
from distant light sources that are at rest in the used coordinate system, must be redshifted. A particle existing in empty space is pushed away from the origin with an acceleration that
increases with distance. In de Sitter's world therefore two reasons for the redshift of spectral lines do exist: firstly the general  
 dispersal of matter towards a matter horizon and secondly the spectral shift towards red which for distant objects occurs  due to the slowing down of time, even if they are at rest with respect
 to the origin of coordinates.

Because of the symmetry of de Sitter's world formulae one can choose an arbitrary point as origin; meaning that there is no difference regarding the observed phenomena between 
the origin and the horizon. Any point can be origin, through every point of de Sitter's universe can pass the mass horizon, the unreachable periphery of space.

The conclusion for the stagnancy of time is rejected by de Sitter with the remark that all paradoxes of this world apply only to events that  occur before or after eternity.

If one objects that de Sitter's world would not be static once matter has been brought in, then this property rather speaks in favour of its physical reality than against it.

Is it possible to observe the slowing down of natural processes with increasing distance?

de Sitter pointed already at the confirmed strong redshifts in the spectra of three spiral nebulae which, calculated as a Doppler effect, result in an average velocity of +600 km/s.

De Sitter's theory not only provides the mere fact of redshift, but, as mentioned, also the increase of this redshift with increasing distance and this for two reasons. This must appear
in the observation in the manner that the Doppler  radial motions of the spiral nebulae assume higher and higher positive values.  
The distance of the spiral nebulae are unknown; however, one might use the apparent diameter as a measure of the distance under the assumption of the same linear diameter on the
average. In de Sitter's cosmology the radial velocities should increase with decreasing apparent diameter.

To prove these expectations, a sufficiently large sample of velocities is available that contains objects with large and small apparent diameter. 
 In its majority one owes these measurements, which are as difficult as  valuable, to  the work of V.M. Slipher, Lowell observatory. His values of radial motions
 are collected in a list that stems from February 1921 and which moreoever contains some still unpublished data; it can be found at Eddington l.c. p.162.
 In addition, another source provides the radial motion for NGC 2681 of +700 km/s, so that velocities exist for 42 spiral nebulae.
 The photographic apparent diameters have been partly borrowed from H.D. Curtis, Lick Public. {\bf 13}, P.I (1918), and partly from F.G. Pease, Mt. Wilson Contr. Nr.132 (1917), Nr. 186 (1920).
 
 The following results of some little calculations are based on the nebulae major axes in arcmin, for obvious reasons not linearly, but in log Dm. After rebinning in suitable groups of the
 two arguments log Dm and v, one gets the table  
 

\begin{center}
\begin{table}[h!]%
\centering
\caption{Radial velocities vs. angular diameter and vice versa \label{tab2}}%
\tabcolsep=0pt%
\begin{tabular*}{20pc}{@{\extracolsep\fill}lcc|ccc@{\extracolsep\fill}}
\toprule
 & argum. log Dm &  & & argum. v  &  \\
\toprule
\textbf{log Dm} & \textbf{v}  & \textbf{n} & \textbf{v} & \textbf{log Dm} & \textbf{n} \\
\midrule
0.24 & +827   & 9  & -183   & 1.11 & 5   \\
0.43 & +656   & 7 & +255 & 0.76 & 8 \\
0.66 & +512   &  8 & +475 & 0.96 & 6 \\
0.88 & +555 & 10 &  +630 & 0.52 & 8 \\
1.07 & +334 & 5 & +809 & 0.48 & 7 \\
1.71 & -20   & 3 & +1160 & 0.60 & 8\\
\bottomrule
\end{tabular*}
\end{table}
\end{center}

One clearly recognises a relation in the sense that the positive radial motion decreases with increasing apparent diameter. Less strikingly,
but still distinctively, one sees that with increasing velocity, the apparent diameter decreases. The degree of correlation of the two parameters,
v and log Dm, is expressed by the correlation coefficient that has been determined from 42 pairs
$$ r = -0.455\pm0.122, n=42 $$ 
Mean values: v=+574; log Dm=0.71 (geom. mean=5'.1) with the linear regression lines:
$$v(km) = 914 - 479\times log Dm$$ and $$log Dm = 0.96   - 0.000432\times v$$

There remains no doubt that the positive radial motion of spiral nebulae increases considerably with increasing distance. In reality, the mutual
conditionality between v and log Dm may be stronger than the formal value of r indicates. In a plot of  the individual values v against log Dm, the
resulting distribution has a triangular or V-shape from which one reads off the following facts: among the small nebulae appear lowest and highest v,
the apparently large nebulae have the lowest v, among nebulae with  low v's, one finds small and large objects, large nebulae with high v do not
exist.

Thus it can be concluded that the scatter of the linear nebulae dimensions fills the triangular area of the graph in the way that among the near nebulae with
low v  absolute small and large objects exist and are visible, while in the deep Universe only the absolute largest nebulae with high v are accessible for
determination of their radial motions. 
Then the relation between log Dm and v would be best represented by the hypotenuse of the triangulum, that limits the data points with the consideration
that the giants among the nebulae have on the average the same linear dimensions at all distances. For these biggest nebulae the resulting formulae
for the graphic representation is parenthetically
$$ v(km) = 2200 - 1200\times log Dm $$
 
The coefficient 1200 expresses that a tenfold increase of the distance, the redshift as a pure Doppler effect increases by 1200 km. 
This rule-of-thumb has an advantage: it does not clash with the speed of light, because the true and apparent radial velocities of distant matter that follow from
 the properties of de Sitter's space increase only slowly with distance. A redshift that would mean light velocity as a Doppler effect would only be reached in distances
 that lie beyond those  nowadays believed  to be the distances of heavenly bodies - about $10^{200}$ parsec, a value that exceeds all estimations for  the radius of
 curved space.
 
 For 32 of 42 nebulae with radial velocities, one can find photometric surface brightnesses in Lund Meddel.(2), 29, (1923), and it deserves mentioning in this context
 that the new larger data set gives the same correlation between radial velocity and surface brightness, as in LM, p.31;  a correlation coefficient $r = +0.02 \pm 0.18$,
 i.e.  there is no relation.
 
 However, the positive sign indicates that radial motions increase with decreasing surface brightness. This sense one finds in all correlations that have been searched for
 where the distance of these entities is somehow involved. Regarding the absolute values at the borderline of reality, but regarding the sign always indicating a (very tiny)
 absorption in the depths of the universe to which one pushes the nebulae. Compare AN 222.33 (1924).
 
 The relation between the radial motions of the nebulae and their distances has been recognised in another way  already in AN 215, p. 352 (1922). It was shown that
 the intrinsic radial motions of the nebulae (with their signs) increase with increasing magnitudes of their total brightnesses. This relation is characterised by a correlation
 $r = +0.21$ from 19 objects, while here from 42 objects the correlation coefficient radial motion vs. log(app. Dm) resulted as r = -0.46. A phenomenon that  can be described
 in the above manner by the properties of de Sitter's world.
}



  


 

\bibliography{wirtz_V2}
\end{document}